\documentstyle[Sprocl]{article}      
\def\Journal#1#2#3#4{{#1} {\bf #2}, #3 (#4)}

\def\NPB{{\em Nucl. Phys.} B}
\def\PLB{{\em Phys. Lett.}  B}
\def\PRL{\em Phys. Rev. Lett.}
\def\PRD{{\em Phys. Rev.} D}

\def\be{\begin{equation}}
\def\ee{\end{equation}}
\def\bea{\begin{eqnarray}}
\def\eea{\end{eqnarray}}

\newcommand{\gsim}{\mathrel{\hbox{\rlap{\lower.55ex \hbox {$\sim$}}            
           \kern-.3em \raise.4ex \hbox{$>$}}}}
\newcommand{\lsim}{\mathrel{\hbox{\rlap{\lower.55ex \hbox {$\sim$}}            
           \kern-.3em \raise.4ex \hbox{$<$}}}}

\newcommand {\beq} {\begin{equation}}
\newcommand {\eeq} {\end{equation}}

\newcommand {\sm}   {standard model}
\newcommand {\ew}   {electroweak}
\newcommand {\ewsm} {electroweak standard model}
\newcommand {\ewpt} {electroweak phase transition}

\newcommand {\YMHth} {Yang-Mills-Higgs theory}
\newcommand {\sph} {sphaleron}

\newcommand {\cs} {configuration space}
\newcommand {\Cs} {Configuration space}

\newcommand {\ES} {E_{{\rm S}}}
\newcommand {\Mvac} {M_{\rm vac}}
\newcommand {\Nfam} {N_{\rm fam}}

\newcommand{\D}{\partial}

\def\R {{\rm I\kern-.15em R}}
\def\L{{\rm I\kern-.25em L}}

\def\id{{\rm 1\kern-.12em
        \rule{0.3pt}{1.5ex}\raisebox{0.0ex}{\rule{0.1em}{0.3pt}}}}
\def\JPA{{\em J. Phys. A : Math. \& Gen.}}

\begin{document}
\renewcommand{\thefootnote}{\fnsymbol{footnote}}
\noindent hep-ph/9612386   \hspace*{\fill} KA--TP--26--1996
\vspace*{1\baselineskip}\newline
\title{ELECTROWEAK DEFECTS 
\footnote{ Expanded version of a paper published in
H. Klapdor-Kleingrothaus and Y. Ramachers (eds.),
\emph{Dark Matter in Astro- and Particle Physics},
World Scientific, 1997.}
                           }
\author{ F. R. KLINKHAMER}
\address{ Institut f{\"u}r Theoretische Physik, Universit{\"a}t Karlsruhe,\\
 D--76128 Karlsruhe, Germany\\
 E-mail: frans.klinkhamer@physik.uni-karlsruhe.de }
\maketitle
\abstracts{A brief, non-technical review is given of certain
defect-like configurations in the \ewsm, which may have played
an important role in the physics of the early universe.  }
\renewcommand{\thefootnote}{\alph{footnote}}
\setcounter{footnote}{0}

\noindent The history of the early universe is, for the most part, rather
uneventful: the universe expands and cools down. There are, however,
brief moments when something interesting does happen. These are, in reverse
order, the 
epochs of
\begin{itemize}
  \item recombination at temperature $T$ = O(eV),
  \item nucleosynthesis at $T$ = O(MeV),
  \item quark confinement at $T$ = O(GeV),
  \item electroweak phase transition at $T$ = O(TeV),
\end{itemize}
where the temperatures indicate the physics involved, namely the physics
of the atomic, nuclear, chromodynamic and weak interactions
(units are such that $\hbar = c = k =1$). 
What happened at even higher
temperatures is a complete mystery, because we do not know how the
elementary particles behave at collision energies significantly above
$10^{12} \, {\rm eV}$ $\equiv$ $1 \, {\rm TeV}$.
Still, it is conceivable that already the \ewpt ~epoch may have some bearing
on the dark matter problem (the nature and origin
of what constitutes the bulk of our present universe).
We therefore present at this workshop some results on an
unusual, but interesting, aspect of electroweak physics.
\vspace*{1\baselineskip}\newline
Elementary particle physics, over the years, has established a so-called
\sm. It consists of two parts: the \ew ~interactions and the strong
(chromodynamic) interactions. \emph{A priori\/}, these two interactions seem to be
unrelated. The most intriguing part of the \sm~is the \ew ~sector, with
the violation of parity (P) and charge conjugation (C)
discovered in the 1950's. The non-conservation of these two discrete
symmetries is nowadays incorporated into the \ewsm, right from the start.
In fact, the non-conservation already shows up in the fields present and
their basic interaction structure (e.g. the absence of an interacting
right-handed neutrino field).
\par This brings us to the main point of our talk, namely that the \ewsm ~
has been established, so far, in perturbation theory only.
The Feynman diagram of Fig. 1, for example, contributes to the
scattering of two electrons ($e^{-}$) by the exchange of a virtual photon
($\gamma$)
or neutral vector boson ($Z$). Essentially, the electrons are
plane waves (wave packets, rather), which now and then emit
or absorb a photon, with a probability amplitude proportional to
the electric charge $e$ of the electron. Standard perturbation theory is the
collection of all relevant Feynman diagrams, which can be ordered as an infinite
series in powers of $e^2$ for the probabilities (Fig. 1, for example,
contributes in order $e^4$ to the scattering cross-section).
But there may be more to the \ew ~theory than just
Feynman diagrams. Indeed, we now believe that there are certain
non-perturbative processes, proportional to $\exp[\,-\,1/e^{2}\,]$,
which may lead to the violation of time reversal symmetry (T) and fermion
number conservation (B + L, with B the baryon and L the lepton
number operator). Remark that the probability amplitude $\exp[\,-\,1/e^{2}\,]$ has
vanishing Taylor expansion around $e=0$, so that there are no corresponding
Feynman diagrams.
For these processes there is no longer a meaningful
distinction between a free and an interacting part of the Hamiltonian,
all terms being equally important. The field configurations involved
are thus very different from plane waves (free particles) and
generically go under the name of ``defects''. Let us first discuss two such
defects in simple models and then see wether or not defect-like
configurations also appear in the \ewsm.
\begin{figure}
\begin{center}
\unitlength1cm
\begin{picture}(6,3)
\put(0,-2.5){\includegraphics{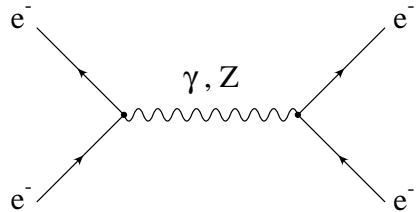}}
\end{picture}
\end{center}
\caption{Feynman diagram.}
\end{figure}
%
\vspace{1\baselineskip}\newline
Consider the Abelian Higgs model in 3+1 dimensions. The cartesian coordinates
are written as ($x^1,x^2,x^3,x^0$) and, for later use, we define the
cylindrical coordinates $\rho$, $\varphi$, $z$ and $t$  by setting
$(\rho \cos\varphi, \rho \sin\varphi, z, t) = (x^1,x^2,x^3,x^0)$.
Static (time-independent) fields have an energy density
\beq
\epsilon =
          \left.\left. \frac{1}{4}\, \right( \D_k A_l - \D_l A_k \right)^2 +
          \left|\, \D_k\, \Phi + i\, \frac{e}{2}\, A_k\, \Phi \,\right|^2 +
          \lambda \left( \Phi^{\star} \Phi - \frac{v^2}{2} \right)^2 \; ,
\label{eq:edens}
\eeq
with indices $k$, $l$ running over 1, 2, 3, and $\D_k$ standing for the partial
derivative $\D/\D x^k$.
Here, $A_k$ is the real gauge field and $\Phi = \phi_1 + i\phi_2$
the complex scalar Higgs field,
with vacuum expectation value $v/\sqrt{2}$,
quartic coupling constant $\lambda$
and electric charge $\frac{1}{2}\,e$. The theory has an $U(1)$ gauge invariance
\begin{eqnarray}
\Phi(x) & \rightarrow & e^{i \omega(x)}\, \Phi(x)             \; ,\nonumber \\
A_k(x)  & \rightarrow & A_k(x) - \frac{2}{e}\, \D_k\, \omega(x) \; ,
\label{eq:U1gaugetransf}
\end{eqnarray}
with $\omega(x)$ an arbitrary real function, so that
$e^{i \omega(x)} \in U(1)$.
The $U(1)$ gauge symmetry group is called Abelian, because a successive
transformation with first $\omega_1$ and then $\omega_2$ gives
the same result as with the order reversed (i. e. the group addition is
commutative). The so-called Higgs symmetry breaking mechanism transforms the
gauge field $A$ into a massive vector field, with a vector boson mass given
by $M_A = \frac{1}{2}\, e\, v$.

As a further restriction we impose $z$-independence, so that we
have an essentially 2-dimensional theory with fields $\Phi(x^1,x^2)$
and $A_k(x^1,x^2)$ for $k=1$, $2$ .
Field configurations with finite string tension
(total energy per unit length in the $z$-direction) obey the following
condition:
\beq
\lim_{\rho \rightarrow \infty} |\Phi| \equiv |\Phi^{\infty}|
                                      =      \frac{v}{\sqrt{2}} \; ,
\label{eq:Phibc}
\eeq
otherwise the integrated scalar potential (\ref{eq:edens})
gives an infinite contribution to the string tension.
The field values $|\Phi|$ $=$ $v/\sqrt{2}$ define the
vacuum manifold $\Mvac$, which has the topology of a circle $S_1$.
Continuous, finite string tension field configurations are thus characterized by their
behaviour at spatial infinity $S_1^{\infty}$ (a limiting circle in the
plane $z=0$, say). In short,  there is the map
\beq
\Phi^{\infty} \;\; : \;\; S_1^{\infty} \rightarrow \Mvac=S_1 \; .
\label{eq:Phiinftymap}
\eeq
This map $\Phi^{\infty}(\varphi) \in \Mvac$ classifies the different field
configurations according
to the number ($n$) of times the complex phase of the field $\Phi$
winds around, for an azimuthal angle $\varphi$ running from $0$ to $2\pi$;
see Fig. 2.

Intuitively, it is clear that any
configuration in the $n=0$ sector can relax to the classical vacuum solution
($A_k=0, \Phi=v/\sqrt{2}$, modulo gauge transformations),
whereas a configuration in
the $n=\pm 1$ sector cannot. One therefore expects the existence of
another classical solution to which $n=1$ configurations may relax.
This is the well-known vortex solution \cite{NO73}
\begin{eqnarray}
\Phi    & = & \bar{f}(\rho) \,e^{i\varphi} \frac{v}{\sqrt{2}} \; ,\nonumber\\
\vec{A} & = & - \frac{2}{e} \,\frac{\bar{g}(\rho)}{\rho}\;\hat{e}_{\varphi} \; ,
\label{eq:NOvortex}
\end{eqnarray}
with $\hat{e}_{\varphi}$ the azimuthal unit vector and
$\bar{f}(\rho)$, $\bar{g}(\rho)$ radial functions
solving the reduced field equations under the boundary conditions
\beq
\bar{f}(0)= \bar{g}(0)= 0 \;\; , \;\;
\bar{f}(\infty)= \bar{g}(\infty)= 1 \; .
\label{eq:NObcs}
\eeq
\begin{figure}
\begin{center}
\unitlength1cm

\begin{picture}(12,5.7) 
\put(0.3,7){\includegraphics{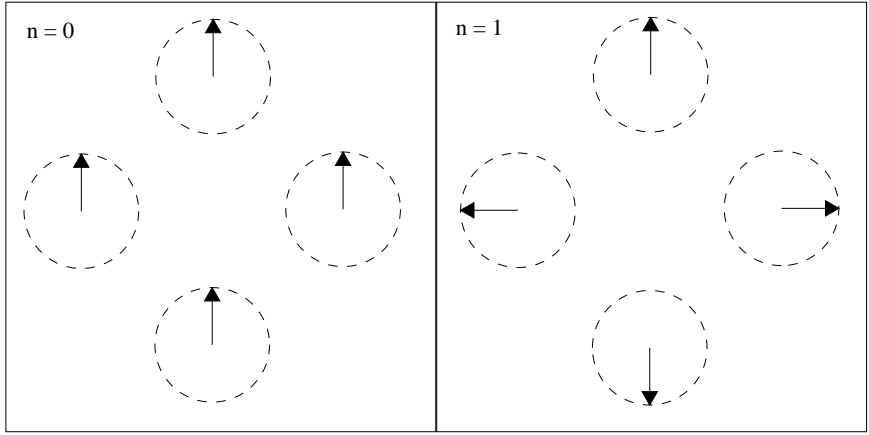}}
\end{picture}
\end{center}
\caption{Winding number sectors.}
\end{figure}
The anti-vortex solution, with winding number $n=-1$, has the signs of
$\varphi$ and $\hat{e}_{\varphi}$ in (\ref{eq:NOvortex}) reversed.
Only the behaviour of the scalar field was sketched in Fig. 2,
but the gauge field plays an equally important role.
The asymptotic gauge field (\ref{eq:NOvortex})
cancels the gradient of $\Phi$ and makes
the first term on the right-hand side of (\ref{eq:edens}) vanish as
$\rho \rightarrow \infty$, thereby keeping the string tension finite.
Moreover, the gauge field
carries a quantized magnetic flux in the $z$-direction.
The solution (\ref{eq:NOvortex}, \ref{eq:NObcs}) is therefore
a flux tube in the $z$-direction, with the Higgs symmetry breaking
absent at the core ($|\Phi| = 0$ for $\rho = 0$),
so that locally the vector field is massless again.

The existence of the vortex solution is governed by the topological
classification $n=\pm 1$, as sketched in Fig. 2.
For this reason, the vortex solution is called a topological defect.
Also, if the highly symmetric vortex configuration (\ref{eq:NOvortex}, \ref{eq:NObcs})  is the
lowest energy classical solution in the $n=1$ sector, then it must necessarily be 
\emph{stable}, since there is no other solution to decay to
(admittedly, the argument is somewhat
circular, but local stability of the vortex solution can be proven rigorously). 
To return to a point made at the beginning,
all terms of the energy density (\ref{eq:edens}) are
of equal importance in the vortex solution 
and there is no separation possible into a free
and an ${\rm O}(e)$ interacting part.
Indeed, the interaction term  $|e\,A_k\,\Phi|^2 $ is $ {\rm O}(1)$, 
due to the large, ${\rm O}(1/e)$, am\-pli\-tu\-de of the vortex
gauge field (\ref{eq:NOvortex}).
All in all, the vortex solution is very different from
the classical vacuum, or small fluctuations around it.

Another interesting example of a topological defect occurs in $SO(3)$
\YMHth. We now consider fully 3-dimensional static
fields and introduce the spherical coordinates $r$, $\theta$
and $\varphi$ by setting
$(r \sin\theta \cos\varphi$, $r \sin\theta \sin\varphi$, $r \cos\theta)$ =
$(x^1,x^2,x^3)$. This theory has a real Higgs triplet
\beq
\Phi = \left( \begin{array}{c}\phi_1\\ \phi_2\\ \phi_3\end{array} \right)\, ,
\label{Higgstriplet}
\eeq
on which a non-Abelian (non-commutative) gauge symmetry operates,
$\Phi(x) \rightarrow \Omega(x) \Phi(x)$ with $\Omega(x) \in SO(3)$.
Finite total energy requires $|\Phi| \rightarrow v/\sqrt{2}$ for
$r \rightarrow \infty$  and
the vacuum manifold has the topology of a sphere, $\Mvac= S_2$.
The relevant map is from the sphere at spatial infinity into
the vacuum manifold
\beq
\Phi^{\infty} \;\; : \;\; S_2^{\infty} \rightarrow \Mvac=S_2 \; .
\label{eq:Psiinftymap}
\eeq
There is now a 3-dimensional generalization of Fig. 2
with winding number $n=1$ (imagine the field vectors pointing outwards,
just as for a frightened hedgehog).
Such a configuration with $n=1$ cannot relax to the vacuum solution,
but may evolve towards another classical solution, the magnetic-monopole
solution.\cite{HP74}
This concludes our brief review of so-called topological
defects. Generally speaking,
these \emph{stable} defects are relevant to the \emph{static}
properties of the theory, for example the spectrum of the
Hamiltonian.
\vspace{1\baselineskip}\newline
We now come to 
the crucial question: can these stable, topological defects be
carried over to the \ewsm ~(EWSM)?
The short answer is: no. The reason is that
the \ew ~theory has a complex Higgs doublet
\beq
\Phi = \left( \begin{array}{c}\phi_1 + i \phi_2\\
                              \phi_3 + i \phi_4 \end{array} \right)\, ,
\label{Higgsdoublet}
\eeq
so that $\Mvac= S_3$. This implies that the mappings from either
the circle $S_1^{\infty}$
or the sphere $S_2^{\infty}$ at infinity into $\Mvac= S_3$
are topologically trivial (in Coleman's words: you cannot lasso
a basketball). Hence, there are no static, stable topological defects
in the $3+1$ dimensional EWSM.
Still (and this is the long answer to the question above),
there are other types of defects in the EWSM, but they turn out to
be \emph{unstable}. It is believed that these unstable defects may be
relevant to the \emph{dynamic} properties of the theory. Here, we will
discuss two such \ew ~defects, the \sph ~and the Z-string.

The existence of the \sph ~solution in the EWSM again follows from
a topological argument. However, rather than considering the
behaviour of the field configurations directly, such as was done in Fig. 2,
one considers instead the topology of \cs. \Cs ~here is the
abstract, mathematical space of 3-dimensional, finite energy configurations
(with the gauge freedom eliminated). Each point of \cs ~corresponds to
a snapshot of the fields. There is, for example, one point in \cs ~which
identifies the vacuum configuration (Higgs field constant and gauge fields vanishing). 
The energy functional $E_{\rm B}$ defines a surface over
configuration space, with the stationary points corresponding to
the solutions of the classical field equations. This \cs ~is
infinite dimensional and non-compact. Moreover, the topology of
\cs ~turns out to be highly non-trivial:
\cs~is not simply an infinite dimensional
euclidian space $\R^{\infty}$.
In fact, there are \emph{holes} and the point at the ``top'' of one such hole
(to be discussed further in the next paragraph)
corresponds to a new classical solution, the so-called sphaleron 
\footnote{From the Greek adjective `sphaleros', meaning ready to fall.} \cite{KM84}.
The \sph ~S, whose explicit field configurations are somewhat complicated,
has typical field energy 
\beq
\ES \equiv E_{\rm B} [\,{\rm S}\,] =
    {\rm O}(M_W / \alpha_w) \sim 10 \, {\rm TeV} \; ,
\label{eq:ES}
\eeq
with $M_W \equiv \frac{1}{2}\, g\,v = 80.4 \,{\rm GeV}$ the mass of the
$W^{\pm}$ vector bosons,
$v = 247 \,{\rm GeV}$ the Higgs vacuum expectation value
and $\alpha_w \equiv g^2 / 4\pi = 1/29.6$ the fine-structure constant
of the $SU(2)$ gauge interactions.
The shape of the \sph ~S is slightly elongated (cigar-like)
and its electromagnetic
field has a large magnetic dipole moment, which may be interpreted \cite{HJ94}
as coming from a tight monopole - antimonopole pair. In this way, we find
magnetic monopoles already in the EWSM!

This particular hole in \cs ~is captured by the following non-trivial map:
\beq
S_1 \times S_2^{\infty} \rightarrow \Mvac=S_3 \; ,
\label{eq:NCLmap}
\eeq
where $S_1$ parametrizes a loop of configurations, each of which is characterized
by its behaviour at spatial infinity $S_2^{\infty}$. 
The \sph ~solution is just one ``point'' on this non-contractible loop in
\cs~(Fig. 3,  left-hand side) and by itself has trivial topology
($S_2^{\infty} \rightarrow S_3$), in contrast to, for example, the single
magnetic monopole solution of the $SO(3)$ theory
based on the non-trivial map (\ref{eq:Psiinftymap}). 
\par Without saying explicitely, we have considered up till now only bosonic
fields (scalar and gauge fields) with total energy $E_{\rm B}$.
We now look at the response of
quantized fermionic fields to these classical bosonic fields.
Specifically, we take a loop of bosonic field configurations around the
configuration space hole mentioned above,
starting and ending at the vacuum V, while passing through the \sph ~S.
This is a non-contractible loop (NCL) of configurations, precisely because
it encircles a hole. If we now consider the Dirac Hamiltonian
eigenvalues ($E_{\rm F}$)
for the different bosonic configurations of the NCL,
we observe the remarkable phenomenon
of \emph{spectral flow}. The NCL starts and ends at the same configuration
(V) and the Dirac eigenvalue spectrum there is the same, but along the
NCL there is a non-trivial flow of eigenvalues (Fig. 3, right-hand side).
The non-trivial spectral flow, whose origin traces back to the
Atiyah-Singer index theorem, has also been observed in
numerical calculations.\cite{KB93}
In the \ew ~context this spectral flow  leads to the violation
of fermion number conservation
\footnote{Note that the Dirac sea (the set of filled negative energy states of the
Dirac Hamiltonian) is truly bottomless, which allows for an overall shift 
of the occupied states as in Fig. 3.} \cite{H76}
\begin{figure}
\unitlength1cm

\begin{picture}(5.0,5.5)
\put(-0.8,5.55){\includegraphics{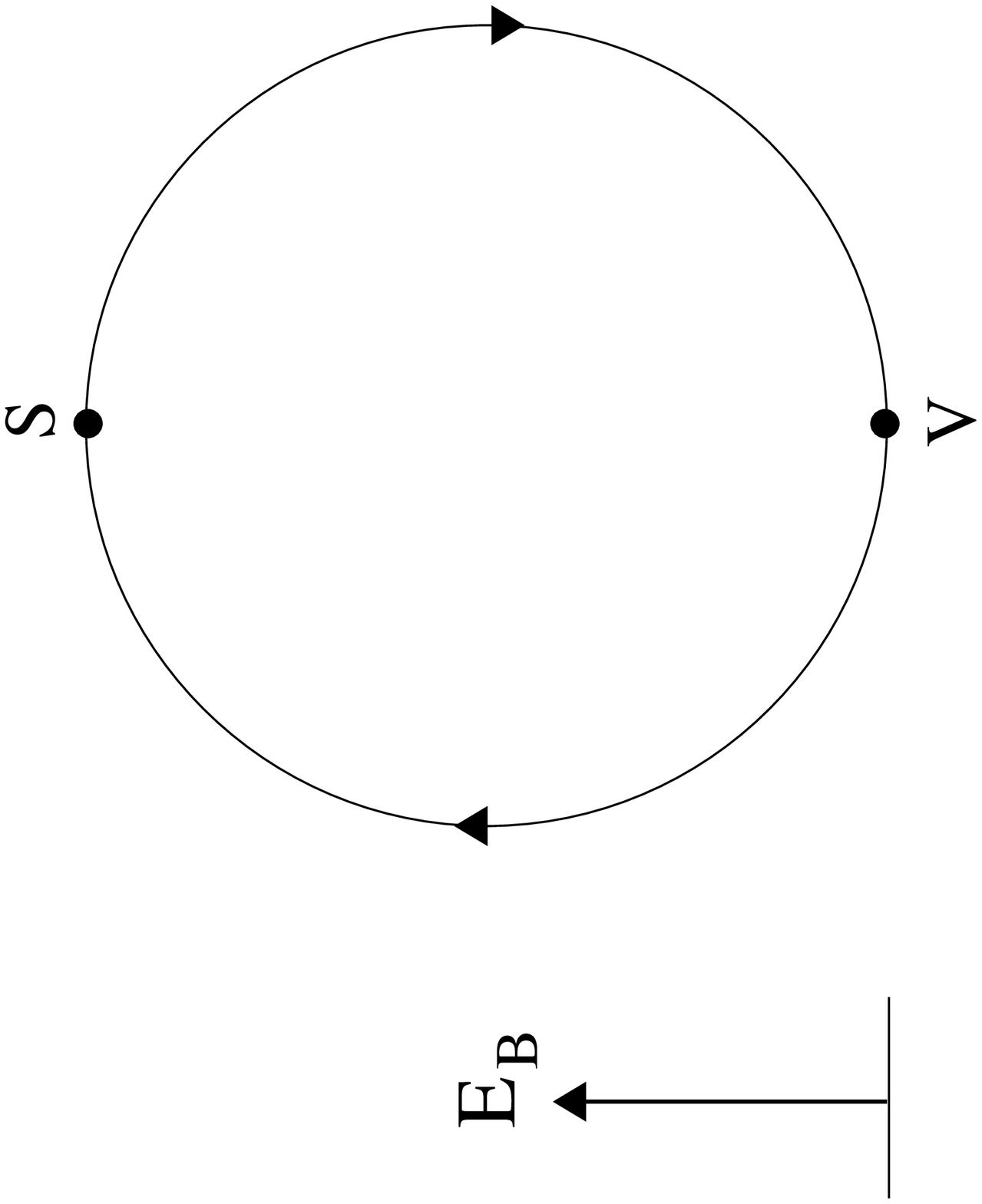}}
\end{picture}
\hfill
\unitlength1cm

\begin{picture}(5.5,5.5)
\put(-0.8,5.55){\includegraphics{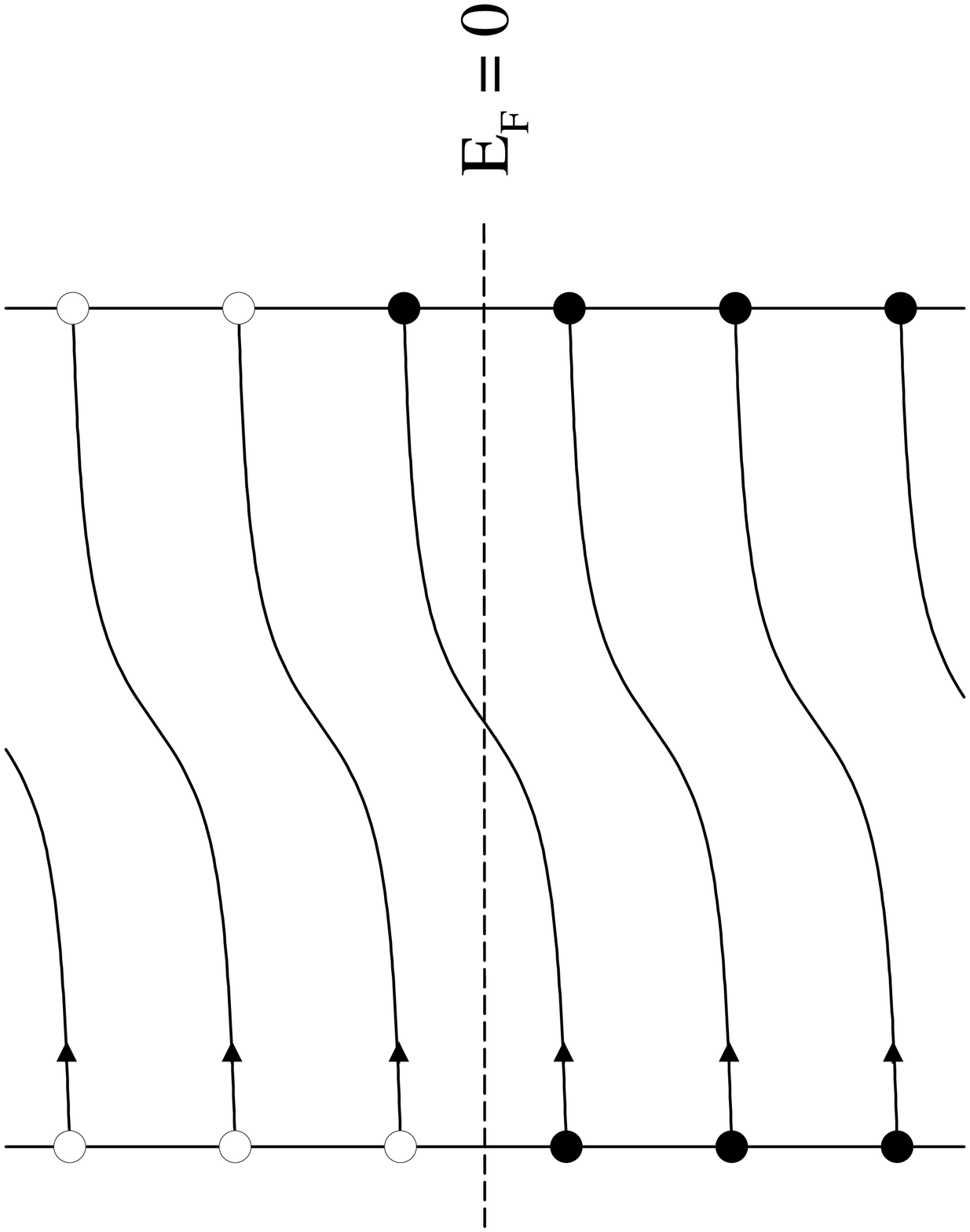}}
\end{picture}
\caption{Non-contractible loop and spectral flow.}
\end{figure}
\beq
\Delta (B + L) = \pm \, 2\, \Nfam \; ,
\label{eq:Delta}
\eeq
where $\Nfam (=3 \:?)$ is the number of families
of quarks and leptons.
The selection rule (\ref{eq:Delta}) is simply the total contribution
of all left-handed fermion doublets (quarks and leptons) 
responding to the $SU(2)$ gauge fields of the NCL.

Originally,  B + L violation was considered \cite{H76}  to proceed
via quantum tunneling (instantons), but the picture of Fig. 3
suggests the possibility of B + L violation at large
temperatures or collision energies, where one passes 
\emph{over} the energy barrier $\ES$, not through. As expected,
 the tunneling rate is
extremely small, with a WKB factor $\exp[\,- 16 \pi^2 / g^2 \hbar\,] \sim
\exp[\, - 372\,]$, whereas the thermal rate, with a Boltzmann factor
$\exp[\,- \ES / k T\,]$, may become significant for large enough temperatures.
These thermal reactions turn out to have been important
in the very early universe.\cite{KM84,KRS85}
However, the precise mechanism for the
\emph{creation} of the present baryon number of the universe is not known,
even though the necessary ingredients are available,
namely the violation of C, CP, B and thermal equilibrium.
This remains one of the outstanding problems of
cosmology and particle physics.

Another classical solution in the EWSM is the Z-string \cite{N77},
which is simply the embedded vortex solution
(\ref{eq:NOvortex}, \ref{eq:NObcs}),
\begin{figure}
\unitlength1cm

\begin{picture}(5.0,5.5)
\put(-0.8,5.55){\includegraphics{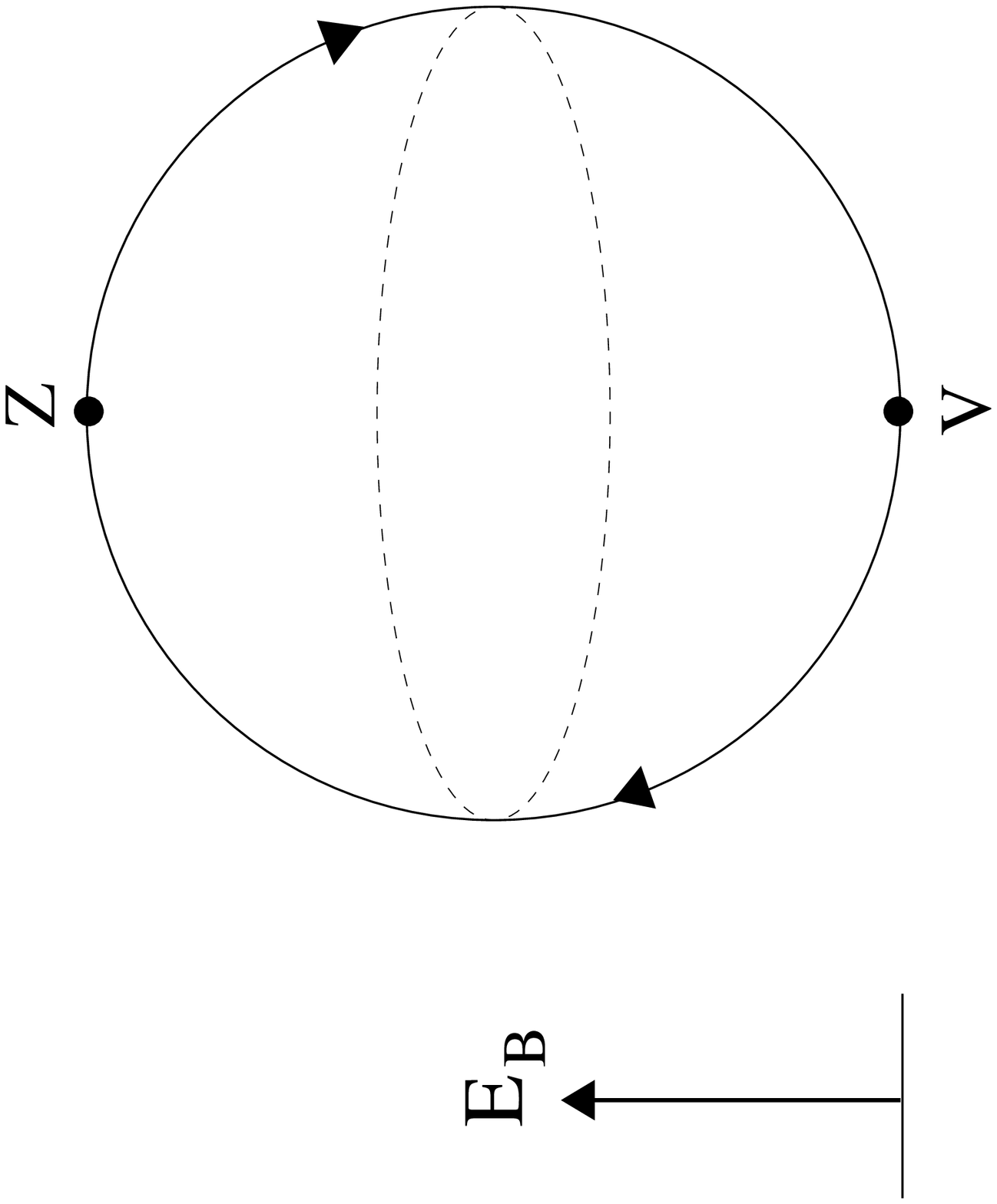}}
\end{picture}
\hfill
\unitlength1cm

\begin{picture}(5.5,5.5)
\put(-0.8,5.55){\includegraphics{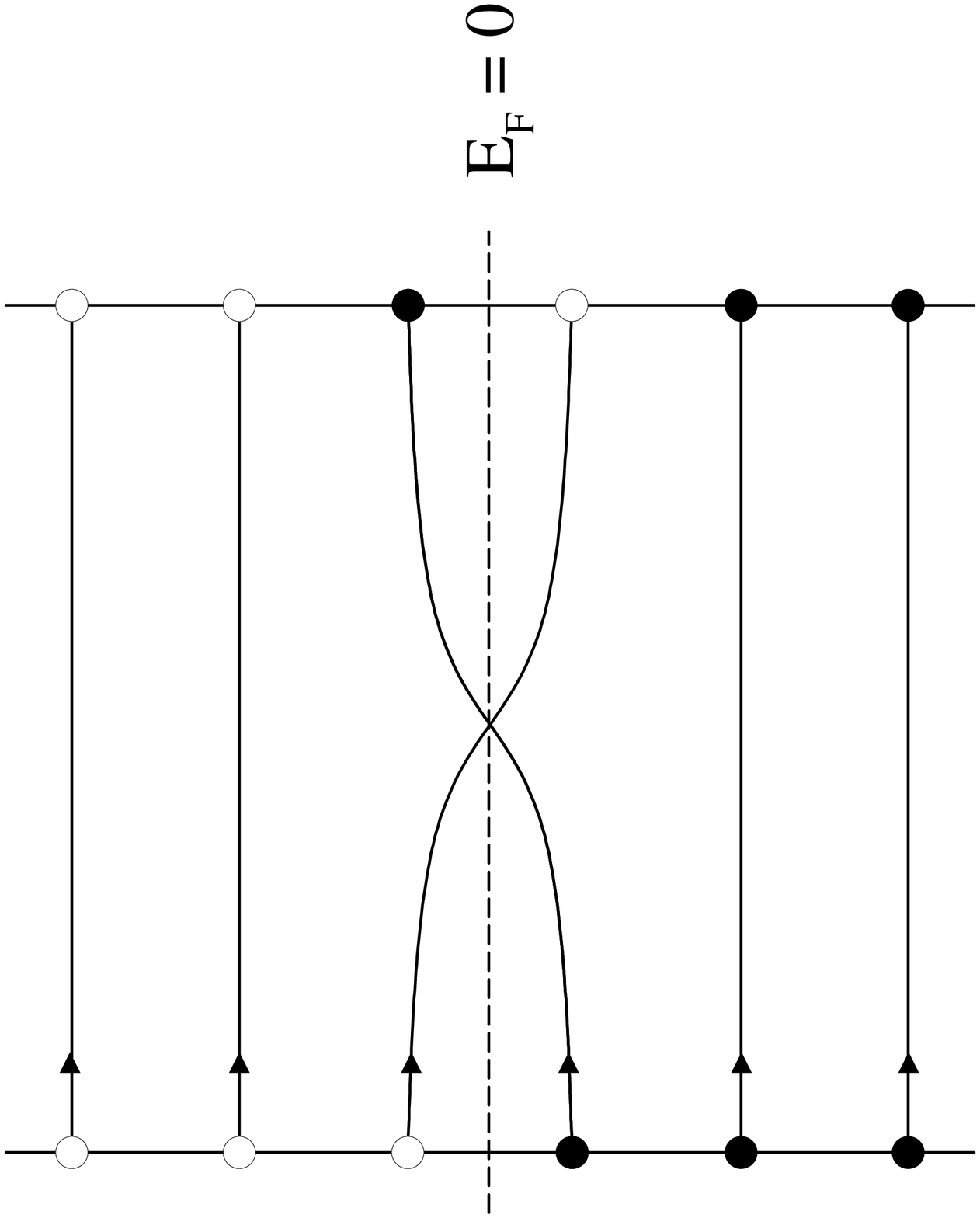}}
\end{picture}
\caption{Non-contractible sphere and level crossing along a meridian.}
\end{figure}
\begin{eqnarray}
\Phi & = & \bar{f}(\rho)
           \left( \begin{array}{c} 0\\e^{i\varphi} \end{array}\right)
           \frac{v}{\sqrt{2}} \; ,\nonumber\\
\vec{Z}  & = & - \,\frac{2\cos\theta_w}{g}\,\frac{\bar{g}(\rho)}{\rho}\,
               \hat{e}_{\varphi} \; ,\nonumber\\
\vec{W}^{\pm} &=& 0 \nonumber \; , \\
\vec{A}       &=& 0           \; ,
\label{eq:Zstring}
\end{eqnarray}
\noindent with $A$ the massless photon field and $Z$, $W^{\pm}$
the massive vector boson fields (mass ratio $M_W / M_Z = \cos\theta_w$,
in terms of the weak mixing angle $\theta_w = 0.491$).
The extra degrees of freedom of the EWSM
turn the Z-string into an unstable solution.
In fact, it can be shown \cite{KO94} that there is a hole in the space of
$z$-independent configurations, around which a non-contractible \emph{sphere}
(NCS) may be constructed, with the Z-string at the top (Fig. 4, left hand side,
where $E_{\rm B}$  indicates the 2-dimensional ``energy'',
i. e. the string tension in the 3-dimensional context).
The instability of the Z-string is ma\-ni\-fest, with at least
two unstable directions. The topologically non-tri\-vi\-al map is now
\beq
S_2 \times S_1^{\infty} \rightarrow \Mvac=S_3 \; ,
\label{eq:NCSmap}
\eeq
where $S_2$ parametrizes a sphere of configurations, each of which is characterized
by its behaviour at spatial infinity $S_1^{\infty}$. 
Again, there is spectral flow of the Dirac
eigenvalues over the NCS.\cite{KR97}
The fermionic levels over the NCS display a cone-like structure and
the behaviour along an arbitrary meridian is sketched in Fig. 4, right-
hand side.
In this case there is no net change of quantum numbers, since fermions
and anti-fermions are created simultaneously.
Still, the non-trivial spectral flow over the NCS may have
implications for the consistency of the theory.\cite{KR97}
\vspace{1\baselineskip}\newline
To conclude, we mention an interesting observation,
which relates these two \ew ~solutions.
The observation \cite{VF94}  is that
\emph{linked} Z-string loops and \emph{twisted}
Z-string segments \footnote{
These particular configurations are, strictly speaking,
not exact static solutions of the field equations and evolve
with time.} (Fig. 5)
have non-vanishing Chern-Simons number, just as the
sphaleron S, and may therefore lead to violation of B + L
(wether or not this actually occurs needs to be verified, though).
This ties in with our earlier remark that the \sph ~solution S can be viewed
as a bound state of a monopole and an antimonopole,
which are then connected by a very short segment of twisted Z-string. 

\begin{figure}
\unitlength1cm

\begin{picture}(4,3.5) 
\put(-1.5,4.55){\includegraphics{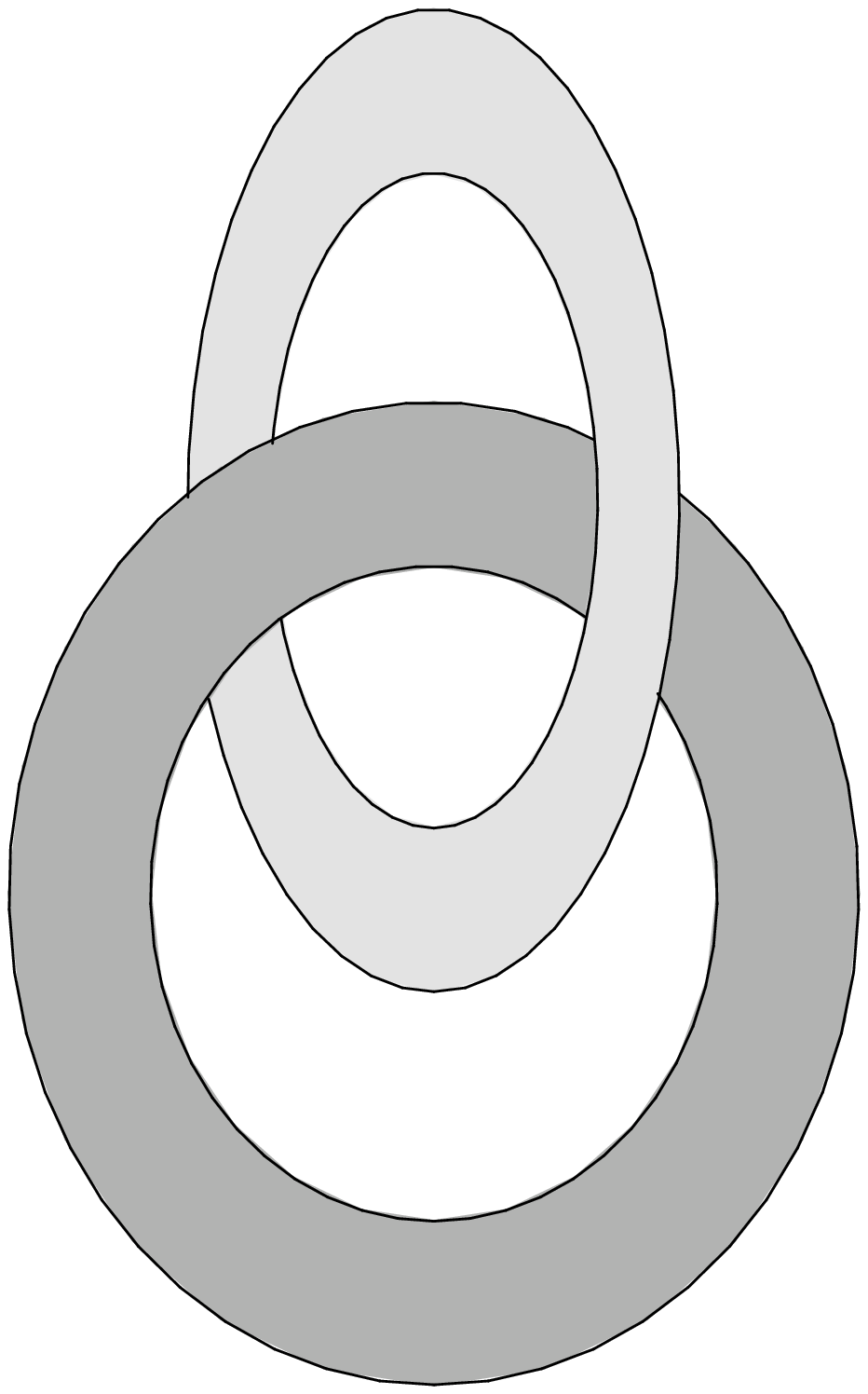}}
\end{picture}
\unitlength1cm

\begin{picture}(4,3.5)
\put(-0.9,4.55){\includegraphics{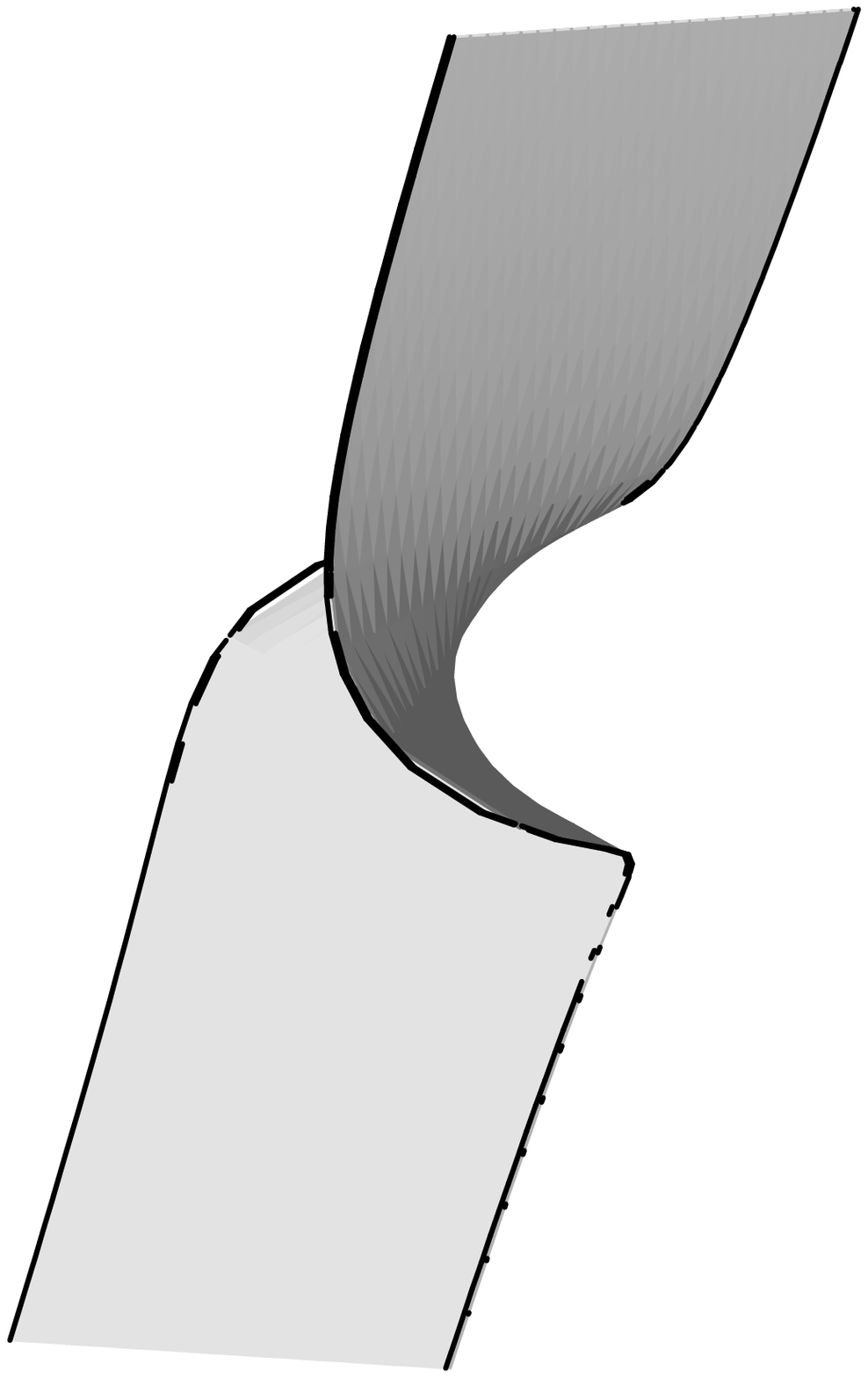}}
\end{picture}

\begin{picture}(3,3.5)
\put(-1.0,4.55){\includegraphics{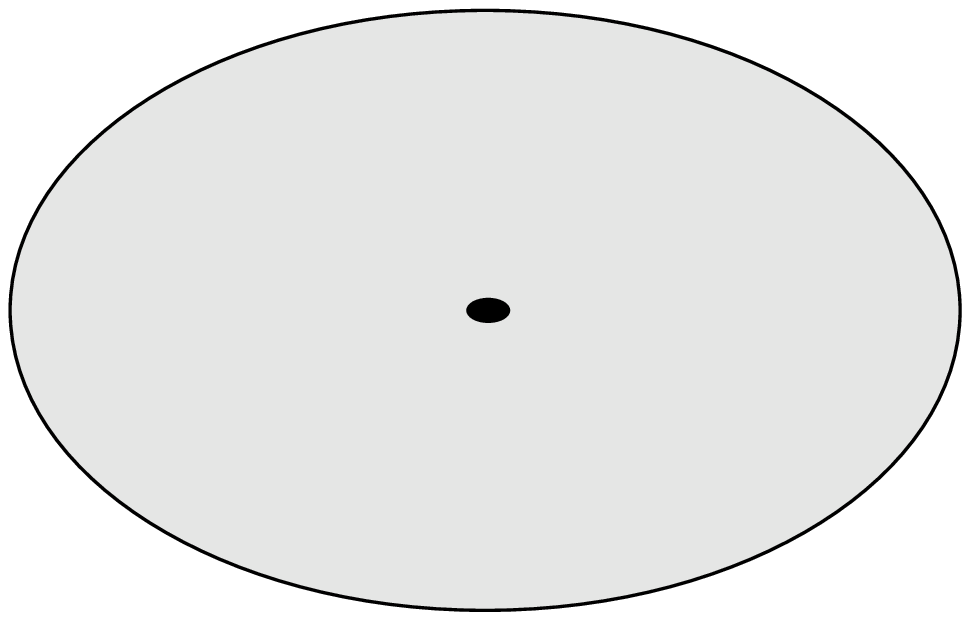}}
\end{picture}
\caption{Linked Z-strings, twisted Z-string segment and \mbox{sphaleron S}.}
\end{figure}

All these unstable defects of the \ewsm~sketched in Fig. 5
are expected to play an important role in the \ewpt ~of the early
universe ($T \sim 100 - 1000 \, {\rm GeV}$).
Extended models of the \ew~and strong interactions
may even have (meta-)stable defects, which could have been created \cite{K76} 
at an earlier phase transition ($T > 1\, {\rm TeV}$). 
Remarkably, there may be the possibility of simulating
the cosmological creation of defects in
appropriate low-temperature condensed matter physics experiments.\cite{Z97}
It is to be hoped that the coming decennia
will provide us with clues, from both elementary particle and
condensed matter physics experiments, to
help explore the cosmological epoch of the \ewpt.


\section*{References}

\begin{thebibliography}{99}
\bibitem{NO73}  H. Nielsen and P. Olesen, \Journal{\NPB}{61}{45}{1973}
\bibitem{HP74}  G. 't Hooft, \Journal{\NPB}{79}{276}{1974};\newline
                A. Polyakov, \Journal{\em Sov. Phys. JETP Lett.\/}{20}{194}{1974}
\bibitem{KM84}  F. Klinkhamer and N. Manton, \Journal{\PRD}{30}{2212}{1984}
\bibitem{HJ94}  M. Hindmarsh and M. James,   \Journal{\PRD}{49}{6109}{1994}
\bibitem{KB93}  J. Kunz and Y. Brihaye,      \Journal{\PLB}{304}{141}{1993}
\bibitem{H76}   G. 't Hooft, \Journal{\PRL}{37}{8}{1976}
\bibitem{KRS85} V. Kuzmin, V. Rubakov and M. Shaposhnikov,
                \Journal{\PLB}{155}{36}{1985};\newline
                P. Arnold and L. McLerran, \Journal{\PRD}{36}{581}{1987};\newline
                M. Dine, O. Lechtenfeld, B. Sakita, W. Fisschler and
                J. Polchinski, \Journal{\NPB}{342}{381}{1990};\newline
                D. Diakonov, M. Polyakov, P. Sieber, J. Schaldach and K. Goeke,
                \Journal{\PRD}{53}{3366}{1996}
\bibitem{N77}   Y. Nambu,      \Journal{\NPB}{130}{505}{1977};\newline
                T. Vachaspati, \Journal{\NPB}{397}{648}{1993}
\bibitem{KO94}  F. Klinkhamer and P. Olesen, \Journal{\NPB}{422}{227}{1994}
\bibitem{KR97}  F. Klinkhamer and C. Rupp, \Journal{\NPB}{495}{172}{1997}
\bibitem{VF94}  T. Vachaspati and G. Field,
                \emph{Phys. Rev. Lett.} {\bf 73}, 373 (1994);
                (E) {\bf 74}, 1258 (1995)
\bibitem{K76}   T. Kibble, \Journal{\JPA}{9}{1387}{1976}
\bibitem{Z97}   W. Zurek, {\em Phys. Rep.\/} {\bf 276}, 177 (1997)
\end{thebibliography}
\end{document}